\documentclass{elsart}
\usepackage{graphics}
\usepackage{graphicx}
\usepackage{epsfig}
\usepackage{amssymb}
\newcommand{\fmn}[2]{\mbox{${\textstyle \frac{#1}{#2}}$}}
\newcommand{\dd}{\mbox{\textrm{d}}}
\newcommand{\reaction}{\mbox{$dp\to\,^{3}\textrm{He}\,\eta$}}

\begin{document}

\begin{frontmatter}

\title{Is there an $\mathbf{\eta\,^3}$He quasi--bound state?}
\author[CW_College]{C.\,Wilkin\corauthref{cor1}},
\ead{cw@hep.ucl.ac.uk} \corauth[cor1]{Corresponding author.}
\author[IKP]{M.\,B\"{u}scher},
\author[IKP,Tbilisi]{D.\,Chiladze},
\author[Dubna]{S.\,Dymov},
\author[IKP]{C.\,Hanhart},
\author[IKP]{M.\,Hartmann},
\author[IKP]{V.\,Hejny},
\author[Tbilisi,Erlangen]{A.\,Kacharava},
\author[IKP,Tbilisi]{I.\,Keshelashvili},
\author[Muenster]{A.\,Khoukaz},
\author[Osaka]{Y.\,Maeda},
\author[Muenster]{T.\,Mersmann},
\author[Muenster]{M.\,Mielke},
\author[Gatchina]{S.\,Mikirtychiants},
\author[Muenster]{M.\,Papenbrock},
\author[IKP]{F.\,Rathmann},
\author[Muenster]{T.\,Rausmann},
\author[IKP]{R.\,Schleichert},
\author[IKP]{H.\,Str\"oher},
\author[Muenster]{A.\,T\"{a}schner},
\author[Gatchina]{Yu.\,Valdau}, and
\author[Cracow1]{A.\,Wro\'{n}ska}
%
%
\address[CW_College]{Physics and Astronomy Department, UCL, London, WC1E 6BT, UK}
\address[IKP]{Institut f\"ur Kernphysik, Forschungszentrum J\"ulich,
  52425 J\"ulich, Germany}
\address[Tbilisi]{High Energy Physics Institute, Tbilisi State University,
0186 Tbilisi, Georgia}
\address[Dubna]{Laboratory of Nuclear Problems, JINR, 141980 Dubna,
Russia}
\address[Erlangen]{Physikalisches Institut II, Universit{\"a}t
Erlangen-N{\"u}rnberg, 91058 Erlangen, Germany}%
\address[Muenster]{Institut f\"ur Kernphysik, Universit\"at
M\"unster, 48149 M\"unster, Germany}%
\address[Osaka]{Research Center for Nuclear Physics, Osaka
University, Ibaraki, Osaka 567-0047, Japan}
\address[Gatchina]{High Energy Physics Department, Petersburg Nuclear
Physics Institute, 188350 Gatchina, Russia}
\address[Cracow1]{Institute of Physics, Jagiellonian University, 30059
Cracow, Poland}
%
%
\begin{abstract}
The observed variation of the total cross section for the
\reaction\ reaction near threshold means that the magnitude of the
$s$--wave amplitude falls very rapidly with the $\eta$
centre--of--mass momentum. It is shown here that recent
measurements of the momentum dependence of the angular
distribution imply a strong variation also in the phase of this
amplitude. Such a behaviour is that expected from a quasi--bound
or virtual $\eta^3$He state. The interpretation can be
investigated further through measurements of the deuteron or
proton analysing powers and/or spin--correlations.
\end{abstract}

\end{frontmatter}
%
%

\begin{center}
\begin{figure}[htb]
\includegraphics[width=14.0cm]{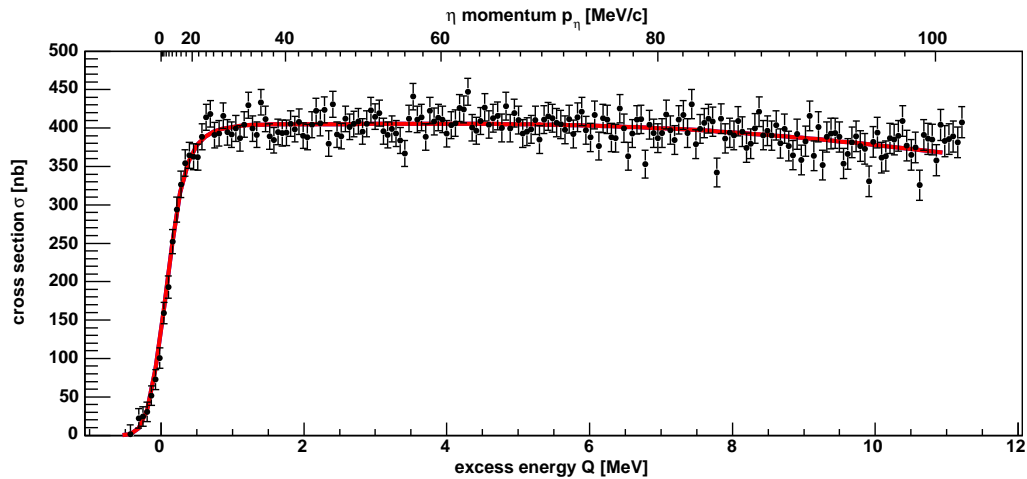}%
\caption{\label{fig:cross} Total cross section for the \reaction\
total cross section measured at COSY--ANKE~\cite{Timo} in terms of
the excess energy $Q$ and $\eta$ c.m.\ momentum $p_{\eta}$. The
fits with and without the $p$--waves, as discussed in the text,
are indistinguishable and so they are not presented separately.}
\end{figure}
\end{center}

New and very precise data on the \reaction\ reaction near
threshold~\cite{Timo,Smyrski}, taken at the COSY accelerator of
the Forschungszentrum J\"ulich, confirm the energy dependence of
the total cross section found in earlier
experiments~\cite{Berger,Mayer}, but with much finer steps in
energy over an extended range. The measurements  at the lowest
excess energy $Q$ (the centre--of--mass kinetic energy in the
$\eta\,^3$He system) are of especial interest. The very rapid rise
and levelling off of the cross section in this region, shown in
Fig.~\ref{fig:cross} for the COSY--ANKE data~\cite{Timo}, suggests
that there is a nearby bound or virtual state of the $\eta\,^3$He
nucleus~\cite{CW,Sib12}.

The concept of $\eta$--mesic nuclei was introduced by Liu and
Haider~\cite{Liu}. Since the $\eta$--meson has isospin--zero, the
attraction noted for the $\eta$--nucleon system should add
coherently when the meson is introduced into a nuclear
environment. On the basis of the rather small $\eta$--nucleon
scattering length $a_{\eta N}$ assumed, they estimated that the
lightest nucleus on which the $\eta$ might bind would be $^{12}$C.
Experimental searches for the signals of such effects have
generally proved negative, as for example in the
$^{16}$O$(\pi^+,p)^{15}$O$^*$ reaction~\cite{Chrien}. The larger
$\textit{Re}(a_{\eta N})$ subsequently advocated~\cite{eta-N}
means that the $\eta$ should bind tightly with such heavy nuclei,
generating large and overlapping widths, and thus be hard to
detect~\cite{later}. On the other hand, it also leads to the
possibility of binding even in light systems, such as
$\eta\,^3$He.

A quasi--bound state leads to a pole of the
$\eta\,^3\textrm{He}\to\eta\,^3\textrm{He}$ scattering amplitude
in the complex momentum $p$ plane with $\textit{Im}(p)>0$ and in
the complex $Q$ plane with $\textit{Im}(Q)<0$%
\footnote{The time evolution of the wave function of the state
involves a factor
$\exp(-iQ_0t)=\exp(-i\textit{Re}(Q_0)t)\exp(+\textit{Im}(Q_0)t)$.
A quasi--bound state must decay in time, which thus requires that
$\textit{Im}(Q_0)<0$. In contrast, a virtual state has
$\textit{Im}(Q_0)>0$ and is also on the second (unphysical) sheet.}. %
Since such a state can decay via the emission of pions or
nucleons, it can only be described as being quasi--bound. If the
$\eta$--nucleus force is not attractive enough, the signs of these
imaginary parts are reversed and the state is called virtual. When
a pole is close to $Q=0$ it distorts strongly the energy
dependence of the \reaction\ total cross section at low energies.
This is precisely what is seen in the experimental
data~\cite{Timo,Smyrski,Berger,Mayer}, with all experiments
identifying a pole with $|Q|$ less than a couple of MeV. The ANKE
data~\cite{Timo} shown in Fig.~\ref{fig:cross} include many points
in the threshold region and, after taking into account the finite
momentum spread of the beam, a pole was identified at
$Q_0=[(-0.30\pm0.15_{\rm{stat}}\pm0.04_{\rm{syst}})\pm
i(0.21\pm0.29_{\rm{stat}}\pm0.06_{\rm{syst}})]$\,MeV, where the
sign of the imaginary part cannot be determined even in principle
from such $\eta$ production data.

The properties of any $\eta\,^3$He nucleus should be largely
independent of the production process but the backgrounds will
be reaction--dependent. The only other evidence for the existence
of the $\eta\,^3$He nucleus has come from
photoproduction~\cite{Pfeiffer}. Though a sharp energy dependence
has been seen in the $\gamma\,^3$He$\to\eta\,^3$He amplitude, the
limited statistics meant that a coarser binning had to be used
than for the \reaction\ reaction~\cite{Timo}. A significant
improvement in this is to be expected from the new MAMI data,
which are currently being analysed~\cite{Krusche}. The MAMI--TAPS group
also found an anomalous behaviour in the photoproduction of
back--to--back $(\pi^-,p)$ pairs. It was suggested that this is
consistent with the existence of a quasi--bound $\eta\,^3$He
state~\cite{Pfeiffer}, though the interpretation is somewhat
controversial~\cite{Hanhart}.

In order to prove that a nearby pole in the complex $Q$ plane is
indeed responsible for the unusual energy dependence of the
\reaction\ cross section, it is necessary to show that the pole
induces a change in the phase as well as in the magnitude of the
$s$--wave amplitude. Since the cross section is proportional to
the absolute square of the amplitude, much phase information is
thereby lost. However, it is the purpose of the present letter to
point out that the interference between the $s$-- and $p$--waves,
as seen in the newly published angular
distributions~\cite{Timo,Smyrski}, leads to the required
confirmation.

\begin{figure}[t!]
\begin{center}
\includegraphics[width=8.0cm]{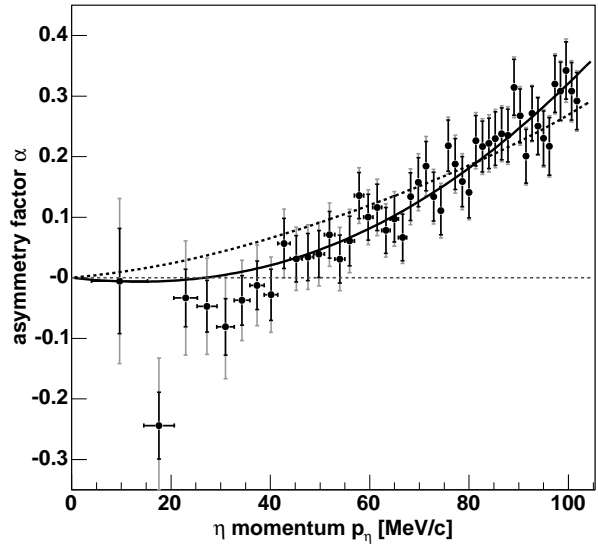}
\caption{\label{angas} Variation of the asymmetry parameter
$\alpha$ defined by Eq.~(\ref{alpha}) with the c.m.\ momentum of
the $\eta$-meson. The solid line is the result of fitting the ANKE
data of Ref.~\cite{Timo} with Eq.~(\ref{alpha3}), where the
$s$--wave amplitude $f_s$ of Eq.~(\ref{eq:fsi2}) was introduced.
The dashed line was obtained after neglecting the phase variation
of $f_s$, i.e.\ letting $f_s\to |f_s|$. In all cases the fits were
carried out by considering only the statistical errors shown by 
dark lines. Combining these quadratically with the systematic
uncertainties leads to the total errors, which are represented 
by the grey lines. }
\end{center}
\end{figure}

The \reaction\ differential cross sections were found to be linear
in $\cos\theta_{\eta}$, where $\theta_{\eta}$ is the c.m.\ angle
between the initial proton and final $\eta$. Throughout the range
of the new COSY measurements, $Q< 11$\,MeV~\cite{Timo,Smyrski},
there is no sign of the $\cos^2\theta_{\eta}$ term that is needed
for the description of the angular distributions at higher
energies~\cite{Bilger}. The angular dependence may therefore be
summarised in terms of an asymmetry parameter $\alpha$, defined as
\begin{equation}
\label{alpha}
\alpha=\left.\frac{\dd\phantom{x}}{\dd(\cos\theta_{\eta})}
\ln\left(\frac{\dd\sigma}{\dd\Omega}\right)\right|_{\cos\theta_{\eta}=0}\:.
\end{equation}
The variation of the ANKE measurements of $\alpha$ with the $\eta$
momentum $p_{\eta}$ is shown in Fig.~\ref{angas}.

On kinematic grounds, the angular dependence near threshold might
be expected to develop like $\vec{p}_p\cdot\vec{p}_{\eta}=
p_pp_{\eta}\cos\theta_{\eta}$, where $\vec{p}_p$ and
$\vec{p}_{\eta}$ are the c.m.\ momenta of the incident proton and
final $\eta$--meson, respectively. However, one striking feature
of Fig.~\ref{angas} is that, although $\alpha$ rises sharply with
$p_{\eta}$, it only does so from about 40\,MeV/c instead of from
the origin, as one might expect on the basis of the above
kinematic argument. At low values of $p_{\eta}$ the error bars are
necessarily large but there seems to be a tendency for $\alpha$
even to go negative in this region. This feature is in line with
the results of other measurements~\cite{Smyrski,Mayer} that have
different systematic uncertainties and so it is likely to be a
genuine effect. Part of this non--linear behaviour arises from the
steep decrease in the magnitude of the $s$--wave amplitude with
momentum. However, the size of the effect observed can only arise
through the rapid variation of the phase of this amplitude, of the
type generated by a nearby pole in the complex $Q$ plane.

There are two independent \reaction\ $s$--wave amplitudes ($A$ and
$B$)~\cite{GW} and five $p$--wave though, to discuss the data
phenomenologically, we retain only the two ($C$ and $D$) that give
a pure $\cos\theta_{\eta}$ dependence in the cross section. The
production operator
\begin{equation}
\label{amps} \hat{f}=A\,{\vec{\varepsilon}}\cdot\hat{p}_p
+iB\,(\vec{\varepsilon}\times\vec{\sigma})\cdot\hat{p}_p
+C\,\vec{\varepsilon}\cdot\vec{p}_{\eta}
+iD\,(\vec{\varepsilon}\times\vec{\sigma})\cdot\vec{p}_{\eta}\,
\end{equation}
has to be sandwiched between $^3$He and proton spinors. Here
$\vec{\varepsilon}$ is the polarisation vector of the deuteron.
The corresponding unpolarised differential cross section depends
upon the spin-averaged value of $|f|^2$
\begin{equation}
\label{cross1} \frac{\dd\sigma}{\dd\Omega}=
\frac{p_{\eta}}{p_p}\,\overline{|f|^2}=\frac{p_{\eta}}{3p_p}\,I\,.
\end{equation}
Using the amplitudes of Eq.~(\ref{amps}) this yields
\begin{equation}
I=|A|^2+2|B|^2+p_{\eta}^{\:2}|C|^2 +2p_{\eta}^{\:2}|D|^2+
2p_{\eta}Re(A^*C+2B^*D)\cos\theta_{\eta}, \label{dsdo2}
\end{equation}
which has the desired linear dependence on $\cos\theta_{\eta}$,
with an asymmetry parameter
\begin{equation}
\alpha=2p_{\eta}\,\frac{Re(A^*C+2B^*D)}
{|A|^2+2|B|^2+p_{\eta}^{\:2}|C|^2+2p_{\eta}^{\:2}|D|^2}\:\cdot
\label{alpha2}
\end{equation}

The strong $\eta^3$He final--state interaction that gives rise to
the quasi--bound pole should affect the two $s$--wave amplitudes
$A$ and $B$ in a similar way and some evidence for this is to be
found from the deuteron tensor analysing power $t_{20}$, which is
small and changes little near threshold~\cite{Berger}. As a
consequence, $|A|\propto |B|$ throughout our range of interest and
it is plausible to represent the data using a spin--average
amplitude.

In the original fit to the whole of the ANKE \reaction\ total
cross section data~\cite{Timo} shown in Fig.~\ref{fig:cross}, any
influence of $p$--waves was neglected and the data represented by
\begin{equation}
f_s = \frac{f_B}{(1-p_\eta/p_1)(1-p_\eta/p_2)}\,, \label{eq:fsi2}
\end{equation}
with
\begin{eqnarray}
\label{mom}
\nonumber%
f_B &=& (50\pm8)\,(\textrm{nb/sr})^{1/2}\,,\\
\nonumber
p_1 &=& [(-5\pm7^{+2}_{-1})\pm i(19\pm2\pm1)]\,\textrm{MeV/c}\,,\\
p_2 &=& [(106\pm5)\pm i(76\pm13^{+1}_{-2})]\,\textrm{MeV/c}\,.
\end{eqnarray}
The first error bar is statistical and the second, where given,
systematic. The error on $f_B$ is dominated by the 15\% luminosity
uncertainty~\cite{Timo}. Note that only the first pole (at
$p_\eta=p_1$) is of physical significance and for this unitarity
requires that $\textit{Re}(p_1)<0$. The signs of the imaginary
parts of the pole positions are not defined by the data. As will
be seen later, the position of the first pole remains stable when
fitting simultaneously the angular dependence and the total cross
section. In contrast, the second pole is introduced to parametrise
the residual energy dependence, which can arise from the reaction
mechanism as well as from a final--state interaction.

Equation~(\ref{eq:fsi2}) shows an $s$--wave amplitude whose phase
and magnitude vary quickly with $p_{\eta}$, but we expect that,
apart from the momentum factor, the $p$--wave amplitudes should be
fairly constant. In the absence of detailed analysing power
information, we take $A=B=f_s$ and $C=D$ to be a complex constant.
With these assumptions the total cross section and
asymmetry parameter become:%
\begin{eqnarray}
\nonumber \sigma&=&\frac{4\pi
p_{\eta}}{p_p}\left[|f_s|^2+p_{\eta}^{\,2}|C|^2\right],\\
\alpha&=&2p_{\eta}\,\frac{Re(f_s^*C)}
{|f_s|^2+p_{\eta}^{\:2}|C|^2}\:\cdot \label{alpha3}
\end{eqnarray}
If the phase variation of the $s$--wave amplitude is neglected, by
replacing $f_s$ by $|f_s|$, the best fit to the asymmetry
parameter does display a little curvature due to the falling of
$|f_s|^2$ with $p_\eta$. Nevertheless, as shown by the dashed line
in Fig.~\ref{angas}, it fails badly to reproduce shape of the
low--momentum data.

On the other hand, when the phase variation of $f_s$ given by
Eq.~(\ref{eq:fsi2}) is retained, the much better description of
the data given by the solid line in Fig.~\ref{angas} is achieved,
with no degradation in the description of the total cross section
presented in Fig.~\ref{fig:cross}. Furthermore, the difference in
the behaviour of $\alpha$ in the low and not--so--low momentum
regions can now be easily understood. The parameters of the fit
are
\begin{eqnarray}
\label{mom2} \nonumber%
f_B &=& (50\pm8)\,(\textrm{nb/sr})^{1/2}\,,\\
\nonumber
C/f_B &=& [(-0.47\pm 0.08\pm0.20) + i(0.33\pm 0.02\pm0.12)]\,(\textrm{GeV/c})^{-1}\,,\\
\nonumber
p_1 &=& [(-4\pm7^{+2}_{-1})- i(19\pm2\pm1)]\,\textrm{MeV/c}\,,\\
p_2 &=& [(103\pm4) -i(74\pm12^{+1}_{-2})]\,\textrm{MeV/c}\,.
\end{eqnarray}

The systematic error in the value of $C$ was estimated by moving
all the points in Fig.~\ref{angas} collectively up and down by one
standard deviation in the systematic uncertainty. Since the
overall phase is unmeasurable, it is permissable to take the $f_B$
of Eq.~(\ref{eq:fsi2}) to be real. Furthermore, because of the
interference between the $s$-- and $p$--waves, the relative phases
of $C$, $p_1$, and $p_2$ do now influence the observables, though
the differential cross section remains unchanged if the signs of
all the imaginary parts are reversed. Compared to the original
solution, where the effects of the $p$--waves were
neglected~\cite{Timo}, the position of the nearby pole $p_1$ is
little changed. This is hardly surprising because this parameter
is mainly fixed by the data from a region which is dominated by
the $s$--waves. Less expected is the very modest change in the
position of $p_2$, which could have been affected more by the
introduction of $C$. As a consequence, the $\eta\,^3$He scattering
length is also changed only marginally to $a=(\pm
10.9+1.0\,i)$\,fm, where the two signs of $\textit{Re}(a)$ again
reflect the possibility of either a quasi--bound or a virtual
state.

The ANKE data indicate that the $s$--wave amplitude for \reaction\
undergoes a very rapid change of phase in the near--threshold
region of the type expected from the presence of a quasi--bound or
virtual $\eta\,^3$He state. When the fits to the COSY-11 results
of Ref.~\cite{Smyrski} are generalised to include the angular
dependence, it is also found that a reasonable description of data
requires that one takes the fast phase variation due to the nearby
pole into account~\cite{Jurek2}.

It is clearly important to try to justify our interpretation
further through the study of other observables, such as the
deuteron and proton analysing powers, which can also be expressed
in terms of the four chosen amplitudes:
\begin{eqnarray}
\nonumber
\sqrt{2}\,I\,t_{20}&=&2\left(|B|^2-|A|^2\right)
+\left(|D|^2-|C|^2\right)p_{\eta}^{\:2}(3\cos^2\theta_{\eta}-1)\\
\nonumber &&+\:4p_{\eta}\cos\theta_{\eta}\,Re\left(B^*D-A^*C\right)\,,\\
\nonumber
I\,t_{21}&=&\sqrt{3}\,\left(Re\left(A^*C-BD^*\right)p_{\eta}\sin\theta_{\eta}
+\:(|C|^2-|D|^2)p_{\eta}^{\:2}\sin\theta_{\eta}\cos\theta_{\eta}\right)\,,\\
\nonumber 2I\,t_{22}&=&\sqrt{3}\,\left(|D|^2-|C|^2\right)p_{\eta}^{\:2}\sin^2\theta_{\eta}\,,\\
\nonumber
I\,it_{11}&=&\sqrt{3}\,\textit{Im}\left(A^*C-BD^*\right)p_{\eta}\sin\theta_{\eta}\,,\\
\nonumber
I\,t_{10}&=&0\,, \\
\label{Ay}
IA_y^{\,p}&=&2\,\textit{Im}\left(A^*D-B^*D+CB^*\right)\,p_{\eta}\sin\theta_{\eta}\,,
\end{eqnarray}
where the $y$ direction is taken along
$\vec{p}_{\eta}\times\vec{p}_p$.

As can be seen from Eq.~(\ref{Ay}), the deuteron spherical
analysing powers $t_{21}$ and $t_{11}$ are sensitive,
respectively, to the real and imaginary parts of an $s$--$p$
interference and another combination is to be found in the
forward/backward asymmetry of $t_{20}$. These will be investigated
in forthcoming experiments at ANKE~\cite{Tobias}. However, if
indeed $A\approx B$ and $C\approx D$, then the magnitudes of the
signals in the polarised deuteron experiments might be small. Even
if this proves to be the case, the proton analysing power $A_y^p$
will not suffer from the same cancellation. This possibility could be
eliminated entirely by measuring the proton analysing power with
an $m=0$ deuteron, because such a tensor spin--correlation
observable is proportional to \textit{Im}$(A^*D)$.

Although we have worked with a restricted number of $p$--wave
amplitudes, the basic elements remain in place when all five are
considered because the analysis depends upon the rapid variation
of the $s$--wave phase compared to the $p$--waves, which are
assumed to have constant phase. Furthermore, it should be noted
that at $\cos\theta_{\eta}=\pm1$ there are indeed only two
$p$--wave amplitudes and so our formulae for the cross section and
$t_{20}$ are exact at these points. The combinations
\begin{eqnarray}
\nonumber \fmn{1}{3}(1-t_{20}\sqrt{2})&=&
\left(|A|^2+p_{\eta}^2|C|^2
\pm 2p_{\eta}\,\textit{Re}(A^*C)\right)\left/ \left(|A|^2+p_{\eta}^2|C|^2\right)\right.,\\
\fmn{1}{3}(1+t_{20}/\sqrt{2})&=& \left(|B|^2+p_{\eta}^2|D|^2\pm
2p_{\eta}\,\textit{Re}(B^*D)\right)\left/\left(|B|^2+p_{\eta}^2|D|^2\right)\right.,
\end{eqnarray}
where the $\pm$ sign refers to forward and backward production,
would then allow one to test the phase variation of $A$ and $B$
separately. Interferences of different nature are to be found in
the spin correlation of transversally polarised protons and vector
polarised deuterons, for which
\begin{equation}
I\,C_{y,y}=-2\textit{Re}\left[A^*B + C^*Dp_{\eta}^{\,2}
\pm(A^*D+BC^*)p_{\eta}\right].
\end{equation}
Experiments to measure both the deuteron tensor analysing powers
and spin--correlations will be undertaken at
COSY--ANKE~\cite{Tobias}.

If the momentum variation of the forward/backward asymmetry in
$\gamma\,^3$He$\to\eta\,^3$He could be measured near
threshold~\cite{Pfeiffer,Krusche}, then this observable should be
influenced by the same $s$--wave phase variation as noted here.
The size of any effect will, of course, depend upon the strengths
of the higher partial waves, and it is possible that these will
enter even faster than for the \reaction\ reaction due to
quasi--free $\eta$ production.

In summary, the angular distribution for the \reaction\ reaction
near threshold is sensitive to an $s$-$p$ interference. The
variation of both the ANKE and COSY-11 experimental data with
$\eta$ momentum requires an extremely strong dependence of both
the phase and magnitude of the $s$--wave production amplitude on
$p_{\eta}$. Such a behaviour is that to be expected from a pole
which is very close to the $\eta\,^3$He threshold, though it is
important to stress that no \reaction\ observable can show whether
this pole lies on the bound or virtual state plane. Because of the
numerous possible decay channels for $_{\eta}^{3}$He, this
distinction has somewhat less significance than that between the
deuteron and the $^1\!S_0$ state of the proton--proton system.

It is reassuring to note that considering the angular distribution
in addition to the total cross section in the fitting procedure
leads to negligible changes in the position of the nearby pole.
This is because it is largely fixed by the very rapid rise in the
cross section close to threshold where the $p$--waves are small.
However, it is the behaviour of the angular distribution which
shows that the interpretation in terms of a pole to be correct.
This belief should be reinforced through the measurement of
\reaction\ analysing powers and spin correlations, which will
allow us to pursue the investigation without some of the
simplifying assumptions which have been made in the current
analysis.

The authors wish to record their thanks to the COSY machine crew
for producing such good experimental conditions and also to other
members of the ANKE collaboration for diverse help in the
experiment. A valuable contribution was made by Yu.N.\,Uzikov. The
support from FFE grants of the J{\"u}lich Research Centre is
gratefully acknowledged.
%
%

\end{document}